# Multi-channel wire gas electron multipliers with gaps between the electrodes of 1 and 3mm

B. M. Ovchinnikov[1,*], V. V. Parusov[1], Yu. B. Ovchinnikov[2]

[1]Institute for Nuclear Research of Russian Academy of Sciences, Moscow, Russia

[2]National Physical Laboratory, Teddington, Middlesex, TW11 0LW, UK

## Abstract

Multi-channel wire gas electron multipliers (MWGEM) with gaps between the electrodes of 1 and 3mm were investigated. The chamber of the MWGEM was filled with pure commercial neon gas at pressure of 0.4 or 1.0 bar and irradiated by $\alpha$-particles $(Pu^{239})$ or $\beta$-particles $(Ni^{63})$. The following maximal coefficients of electrons proportional multiplication have been obtained:

$6 \cdot 10^3$ ($\alpha$, $\delta$=3mm, P=1bar, 20% streamers);

$1.2 \cdot 10^3$ ($\beta^-$, $\delta$=3mm, P=1bar, 50% streamers);

$6 \cdot 10^3$ ($\alpha$, $\delta$=3mm, P=0.4bar, 20% streamers);

$10^5$ ($\beta^-$, $\delta$=3mm, P=0.4bar, 50% streamers).

For the case, when the multiplication took place simultaneously in the MWGEM gap and in its anode gap, the following maximal coefficients of proportional multiplication of electrons have been obtained:

$1.08 \cdot 10^5$ ($\beta^-$, $\delta$=1mm, P=0.4bar, 50% streamers);

$2 \cdot 10^6$ ($\beta^-$, $\delta$=3mm, P=0.4bar, 20% streamers);

$1.12 \cdot 10^5$ ($\alpha$, $\delta$=3mm, P=0.4bar, 50% streamers).

[*]e-mail: ovchin@inr.ru, tel: +8(4967) 51-98-85.

## Introduction

A multi-channel wire gas electron multiplier (MWGEM) has a number of advantages over a standard plastic GEM [1]. In a GEM, the streamer and spark discharges, and also the streams of positive ions from the proportional avalanches, disperse the metallic and resistive coatings of the GEM with subsequent deposition of them on the walls of the GEM openings and on the chamber insulators. The accidental spark and streamer discharges over the plastic surface also create the conductive channels on them. This unavoidably leads to parasitic leakages and discharges of GEM and finally to its complete damage. In addition, the static charges accumulation on the walls of the plastic GEM leads to instability of GEM operation.

These effects do not exist in MWGEM, the space between the electrodes of which is filled with gas. The gaseous ions produced by the sparks, streamers and proportional discharges are quickly removed from the gap between the electrodes of the MWGEM by the electric field potential. On the other hand, the insulators of the MWGEM chamber must be protected from the deposition of conductive coatings.

There are also a number of other advantages of the multi-channel wire gas electron multipliers (particularly in two-phase cryogenic chamber [2]) over plastic gas electron multipliers. First, the spring tension of the wire electrodes of MWGEM can provide their flatness over long period of time, while thin (~0.1 mm) plastic GEM loses their plane form gradually. Second, the MWGEM consists of low-outgassing materials, which reduces the contamination of the working gas of the MWGEM with other gases. Third, as far as MWGEM does not use any solid-state spacers between the electrodes, there is no condensation of saturated steams of xenon (argon) in the gaps of the MWGEM.

## Experimental setup and results

A MWGEM with a gap of 1 mm has been already investigated in works [3, 4]. In this work the operation of the MWGEM of the same design, but with gaps of 1 and 3 mm (Fig. 1), filled with pure commercial neon gas at pressures of 0.4 and 1.0 bar, was investigated. The electrodes of the MWGEM have openings of 0.5×0.5 mm$^2$, with 1.5 mm steps between them in two orthogonal directions, while the total working area has diameter of 20 mm [4]. The gas multiplication of electrons in MWGEM happens between the rectangular openings of the electrodes. The gap between the cathode C and MWGEM (Fig. 1) is equal to 13 mm, and the gap between the anode A and the MWGEM is equal to 6 mm.

The gap between the cathode and MWGEM was irradiated by $\alpha(Pu^{239})$ or $\beta(Ni^{63})$ particles. The signals from the anode were amplified by a charge sensitive amplifier BUS 2-96.

### *MWGEM with a gap 3 mm, P=0.4 and 1.0 bar*

The dependences of the coefficients of proportional electrons multiplication from the potential difference between the MWGEM electrodes, for a gap of 3 mm, are shown in Fig. 2. The field strength in the anode gap for these measurements was less than the threshold of electron multiplication (for P=0.4 bar it was equal to $E_A^{thr}=420$ V·cm$^{-1}$ and for P=1.0 bar it was equal to $E_A^{thr}=500$ V·cm$^{-1}$).

It is observed, that increase of the voltage difference by several volts above critical value, which is corresponding to maximal amplification $K_{ampl}^{max}$, leads increase of the number of streamers up to ~100%, with subsequent transition into continuous current discharge.

### *MWGEM with additional multiplication in anode gap*

To increase the amplification $K_{ampl}$ and the working range, the field strength in the anode gap was increased up to 500 V·cm$^{-1}$. This provided additional amplification $K_{ampl}^{anode}=14$ to the main amplification of MWGEM $K_{ampl}$(MWGEM). As a result, the total coefficient of the multiplication was increased by factor of 14.

The coefficient $K_{ampl}^{anod}=14$ was obtained by irradiation of the MWGEM with $\alpha$-particles and 80 V potential difference between the MWGEM electrodes. This voltage was not strong enough for electron multiplication, but sufficient to transfer electrons through the MWGEM to the anode gap.

The results for the simultaneous MWGEM-Anode proportional amplification ($K_{ampl}^{tot}=K_{ampl}$(MWGEM)$\times K_{ampl}^{anod}$), for the gap widths of 1 and 3 mm and neon pressure of 0.4 bar, are shown in Fig. 3.

### *Discussion of the results*

Comparing the results for the MWGEM without electrons multiplication in the anode gap (Fig. 2), [$\beta^-$, 3 mm, 0.4 bar, $K_{ampl}^{max}$(MWGEM)=$10^5$ (50% streamers)] and

[$\alpha$, 3 mm, 0.4 bar, $K_{ampl}^{max}$(MWGEM)=$6\cdot10^3$ (20% streamers)],

to the MWGEM with the additional electrons multiplication in anode gap (Fig. 3),

[$\beta^-$, 3 mm, 0.4 bar, $K_{tot}=K_{ampl}^{max}$(MWGEM)$\times K_{ampl}^{anod}=1.4\cdot10^5\times14=2\cdot10^6$ (20% streamers)] and

[$\alpha$, 3 mm, 0.4 bar, $K_{tot}=K_{ampl}^{max}$(MWGEM)$\times K_{ampl}^{anod}=8\cdot10^3\times14=1.2\cdot10^5$ (50% streamers)],

it may be concluded, that the additional multiplication in the anode gap does not influence the multiplication of the MWGEM, but increases the range of proportional multiplication (without streamers) by 14 times.

The commercial neon, which was used in this experiment, was contaminated with other gases *($H_2O$, $N_2$, $O_2$)* at concentrations ≤1 ppm. In addition, as far as the chamber of the MWGEM was not baked for cleaning of its walls from the air and water admixtures, the neon in the chamber contained also admixtures of $N_2$, $O_2$ and $H_2O$ at the levels less than 100 ppm. As it was shown earlier in [5], the presence of these admixtures leads to formation of secondary Penning avalanches, which have long duration (20 – 100 μs). On the other hand, the amplitudes of these secondary avalanches, for the concentrations of contaminations in neon at levels ≤100ppm, are small in the whole range of proportional signals and they do not produce streamer discharges.

Only near the maximal coefficient of amplification, $K_{ampl}^{max}$, the secondary avalanches increase and turns into streamers.

The smaller maximal amplification for $\alpha$-particles, compared to $\beta$-particles, $K_{ampl}^{max}(\alpha) < K_{ampl}^{max}(\beta)$, can be explained by production of the secondary avalanches for $\alpha$-particles with larger amplitudes than for $\beta$-particles.

It is already known, that increase of secondary avalanches in mixtures of noble gases with quenching additions (5-20)%, compared to noble gases, containing $(H_2O, N_2, O_2)$ less than 100 ppm leads to decrease of the maximum amplification $K_{ampl}^{max}$:

1. $K_{ampl}^{max}(Ne+5\%CH_4) < K_{ampl}^{max}(Ne)$ [6];

2. It is convincingly shown, that increasing of the quenching additions for the mixtures $Ne+DME$, $Ar+DME$, $Ar$+izobuthylen decreases $K_{ampl}^{max}$ [7];

3. The $K_{ampl}^{max}=100$ for mixture $Ar+20\%CH_4$, but for pure commercial $Ar$ $K_{ampl}^{max}=10^3$ [4].

The quenching additions remove the photon feedback in MWPC. On the other hand, the photon feedback in GEM is small, because of low photon flux to cathode, which is restricted by angles. Therefore, the large quantities of quenching additions, for suppression of the photon feedback, are not necessary in GEM.

The increase of the MWGEM gap from 1 to 3 mm leads to increase the maximal amplification, $K_{ampl}^{max}$, by factor of 10 by expense of increasing of the avalanche multiplication path (see Fig. 3c (1 mm) and Fig. 3b (3 mm)). For the gap 3 mm, the range of proportional multiplication without streamers is also increased up to one order of the magnitude.

## Conclusion

The multi-channel wire gas electron multiplier with a gap of 3mm is produced. The increasing of the MWGEM gap from 1 to 3 mm increase $K_{ampl}^{max}$ and the range of proportional multiplication (without streamers) by one order of the magnitude by expense of increasing of the avalanche multiplication path.

The simultaneous multiplication of electrons in the MWGEM and its anode gap, at neon pressure of 0.4 bar, provides the maximal proportional multiplication coefficient of about $2\cdot10^6$ with the range of proportional multiplication up to $K_{ampl.}=10^6$. The additional multiplication in the anode gap practically does not influence the MWGEM amplification.

From this work and works [4, 5, 6, 7] it can be concluded that the streamers are developed from the secondary Penning avalanches.

The admixture of large quantities (5-20)% of the quenching additions in the noble gases increases the amplitudes of the secondary avalanches and, as a result of this, decreases $K_{ampl}^{max}$ [4, 5, 6, 7].

Finally, the reliable MWGEM can find applications not only in high-energy experimental physics and astronomy, but also in medicine (X-ray examinations positron, tomography) and industrial defectoscopy.

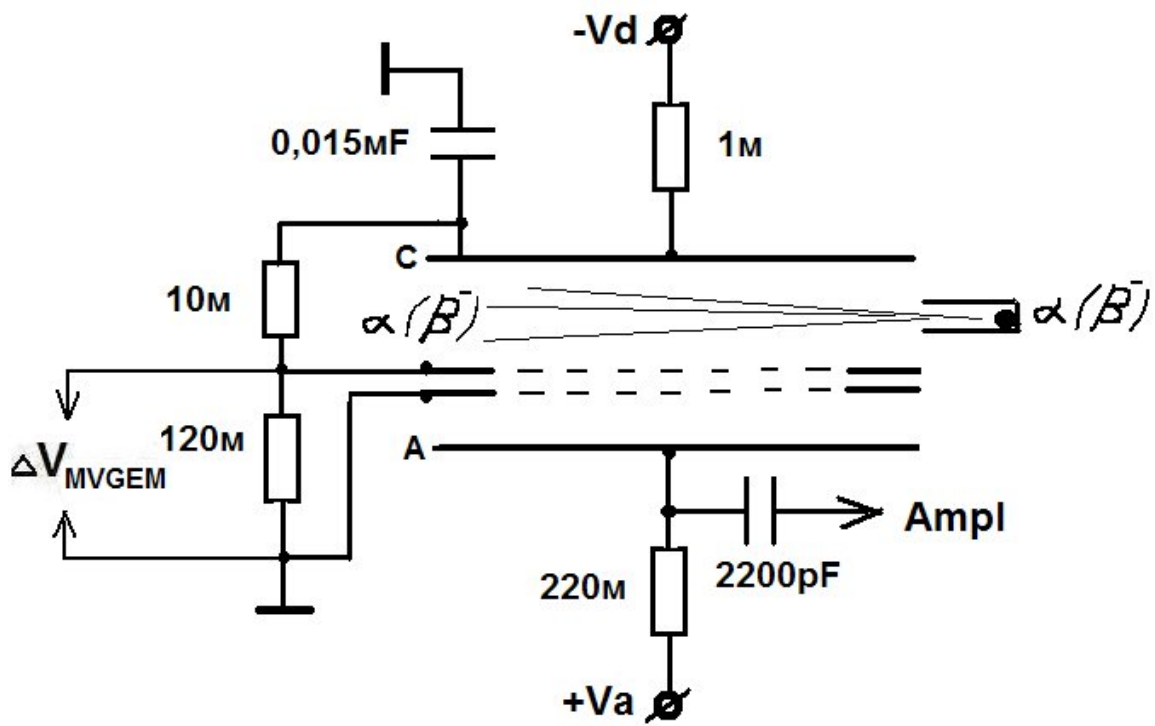

Fig. 1. The chamber for MVGEM investigation.

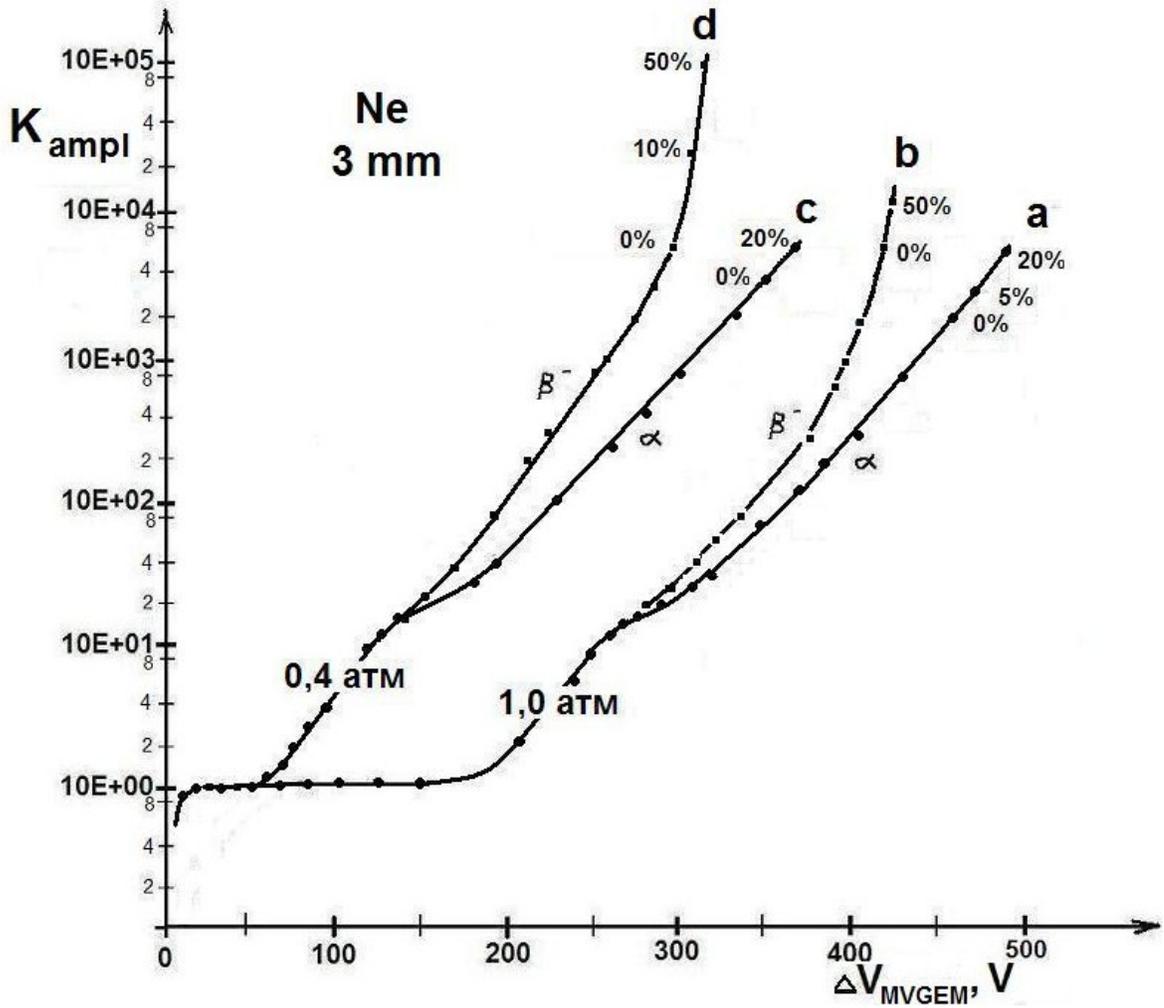

Fig. 2. The dependences of the coefficients of proportional electron multiplication on the potentials difference between MWGEM electrodes for a gap 3 mm at pressures 0.4 and 1.0 bar:

Curve (a): $\alpha$, P=1.0 bar, $K_{ampl}^{max}=6\cdot 10^3$ (20% of streamers), no streamers for $K_{ampl}\leq 2000$;

Curve (b): $\beta^-$, P=1.0 bar, $K_{ampl}^{max}=1.2\cdot 10^4$ (50% of streamers), no streamers for $K_{ampl}\leq 6000$;

Curve (c): $\alpha$, P=0.4 bar, $K_{ampl}^{max}=6\cdot 10^3$ (20% of streamers), no streamers for $K_{ampl}\leq 3.75\cdot 10^3$;

Curve (d): $\beta^-$, P=0.4 bar, $K_{ampl}^{max}=10^5$ (50% of streamers), no streamers for $K_{ampl}\leq 6\cdot 10^3$.

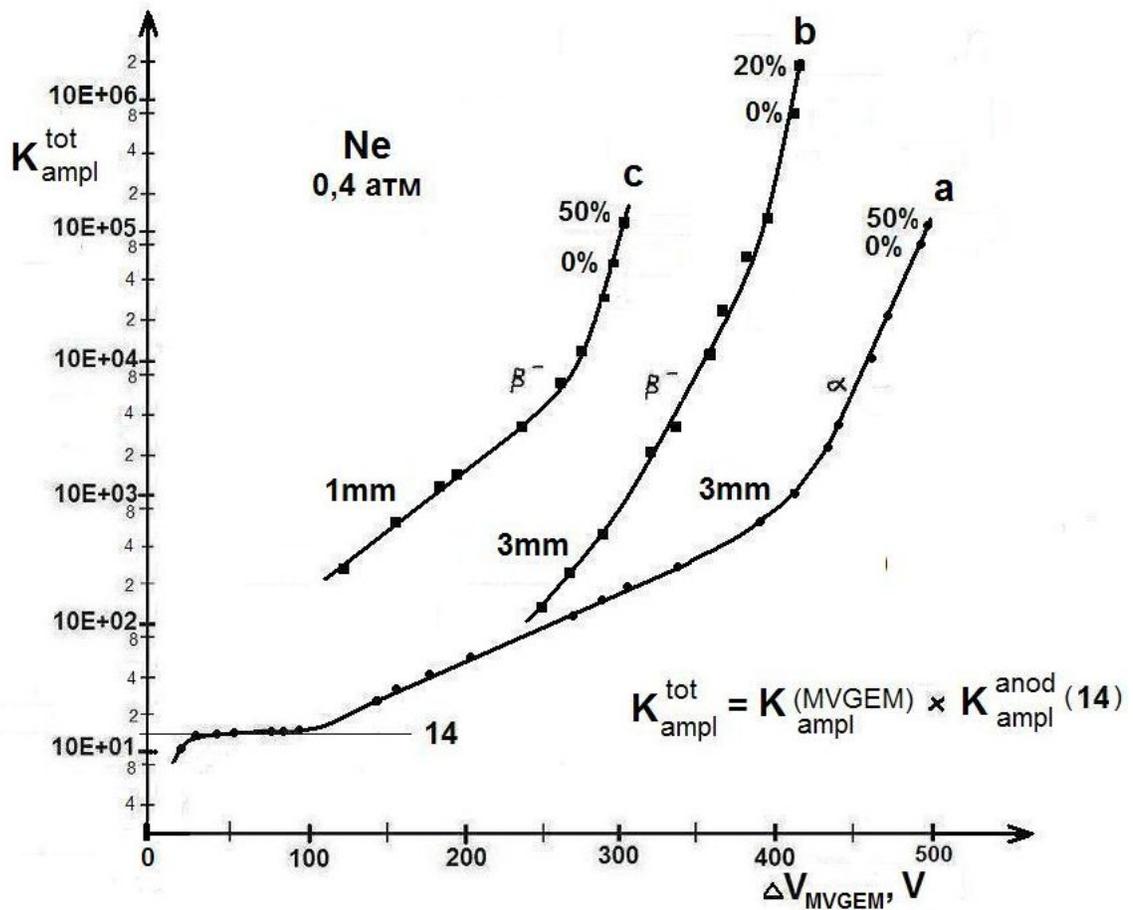

Fig. 3. The dependences of the coefficients of proportional electron multiplication $K_{tot}= K_{ampl}(MWGEM)\times 14$ on the potentials difference between MWGEM electrodes for gaps 1 and 3 mm and neon pressure of 0.4 bar.

Curve (a): $\alpha$, 3 mm, 0.4 bar, $K_{tot}^{max}=1.12\cdot 10^5$ (50% of streamers), no streamers for $K_{ampl}\leq 8\cdot 10^4$;

Curve (b): $\beta^-$, 3 mm, 0.4 bar, $K_{tot}^{max}=2\cdot 10^6$ (20% of streamers), no streamers for $K_{ampl}\leq 7.8\cdot 10^5$;

Curve (c): $\beta^-$, 1 mm, 0.4 bar, $K_{tot}^{max}=1.08\cdot 10^5$ (50% of streamers), no streamers for $K_{ampl}\leq 5.9\cdot 10^4$.